\documentclass[twocolumn,amsmath,showpacs,amsfonts,aps,prc,floatfix]{revtex4}
\usepackage{natbib}
\usepackage{graphicx}
\usepackage{bm}
\begin{document}

\title{Fluctuations in slope parameter in event-by-event hydrodynamics and momentum anisotropy   in heavy ion collisions}

\author{A. K. Chaudhuri}
\email[E-mail:]{akc@vecc.gov.in}
\affiliation{Theoretical Nuclear Physics Group, Variable Energy Cyclotron Centre,
1/AF, Bidhan Nagar, 
Kolkata 700~064, India}

\begin{abstract}

In  event by event hydrodynamic model, we have simulated 30-40\% Au+Au collisions at RHIC and computed the slope parameter from the invariant pion distribution. In each event, the slope parameter fluctuates azimuthally. Fourier expansion coefficients $T_n$ for the slope parameter and the Fourier expansion coefficients $v_n$ for the azimuthal distribution $\frac{dN}{d\phi}$ are found to be strongly correlated. Strong correlation between the two expansion coefficients suggests that in addition to azimuthal distribution, fluctuations in the slope parameter of the invariant distribution can as well be used to study the final state momentum anisotropy in relativistic energy heavy ion collisions.  If measured experimentally, they can serve as additional constraint for hydrodynamical modeling. 
\end{abstract}

\pacs{47.75.+f, 25.75.-q, 25.75.Ld} 

\date{\today}  

\maketitle

\section{Introduction}\label{sec1}

In recent years, there is much interest in event-by-event hydrodynamics. Event-by-event hydrodynamics takes into account that  in  nucleus-nucleus collisions, participant nucleon positions fluctuates from  event to event. Effects of such
  fluctuations are most prominent on 
the azimuthal distribution of the produced particles. In a non-zero impact parameter collision between two identical nuclei, the collision zone is asymmetric. Multiple collisions transform the initial asymmetry   into momentum anisotropy. Momentum anisotropy is best studied by decomposing it   in a Fourier series, 
 
\begin{equation} \label{eq1}
\frac{dN}{d\phi}=\frac{N}{2\pi}\left [1+ 2\sum_n v_n cos(n\phi-n\psi_n)\right ], n=1,2,3...
\end{equation} 
 
\noindent   $\phi$ is the azimuthal angle of the detected particle and 
$\psi_n$ is the  plane of the symmetry of initial collision zone. In central rapidity region, for smooth initial matter distribution (obtained from geometric overlap of density distributions of the colliding nuclei), plane of symmetry of the collision zone coincides with the reaction plane (the plane containing the impact parameter and the beam axis), 
$\psi_n \equiv \Psi_{RP}, \forall n$. The odd Fourier coefficients are zero by symmetry. However, fluctuations in the positions of the participating nucleons can lead to non-smooth density distribution, which will fluctuate on event-by-event basis.  
The participating nucleons then determine the symmetry plane ($\psi_{PP}$), which fluctuate around the reaction plane \cite{Manly:2005zy}. As a result odd harmonics $v_1$, $v_3$, $v_5$ ..., which in central rapidity region are exactly zero for smoothed initial distribution, can be developed. In RHIC and LHC energy collisions, odd harmonics has indeed been measured in experiments \cite{:2011vk}\cite{Adare:2010ux}\cite{Adare:2011tg}\cite{Lacey:2011av}. Explicit event-by-event hydrodynamic simulations of relativistic heavy ion collisions also results in odd harmonics \cite{Schenke:2010rr} \cite{Schenke:2011bn}\cite{Schenke:2012wb}\cite{Gale:2012rq}
\cite{Petersen:2010cw}\cite{Holopainen:2010gz}\cite{Werner:2010aa}\cite{Aguiar:2001ac}
\cite{arXiv:1104.0650}\cite{Bozek:2012fw} \cite{Gardim:2011xv}\cite{Gardim:2012dc}.

While momentum anisotropy is best studied by Fourier expansion of the azimuthal distribution, in the present paper we explore the possibility  of using another observable, namely the slope parameter  of the transverse momentum distribution for the produced particles. In experiments azimuthal angle integrated transverse momentum distribution ($\frac{d^2N}{dyp_Tdp_T}$) of produced particles are routinely measured. Angle integrated momentum  distribution is of exponential nature. Slope ($T$) of the distribution can be interpreted approximately as the temperature of the fireball produced in the collisions. 
In event-by-event hydrodynamics the initial medium is granular and even in central collisions have azimuthal dependence. As a consequence the freeze-out surface can have azimuthal dependence. The invariant particle spectra will depend on the azimuthal angle. 
Hydrodynamic simulations indicate that the invariant spectra $\left (\frac{d^2N}{dyp_Tdp_Td\phi}\right )$ remains exponential even as a function of the azimuth. Asymmetry in the momentum distribution then be reflected as the azimuth dependent slope parameter ($T(\phi)$). Experimental measurements can be easily extended to study of azimuthal fluctuations in the slope parameter. The data can serve as additional constraint for hydrodynamical modeling. 

In the present paper, in event-by-event hydrodynamics, we have simulated $\sqrt{s}_{NN}$=200 GeV 30-40\% Au+Au collisions and studied azimuthal fluctuations of the slope parameter. When averaged over many events,   azimuthal fluctuation in the slope parameter is very similar to that of azimuthal distribution; though less pronounced. In a single event however, the slope parameter appears to fluctuates more strongly over the azimuth than the azimuthal distribution. In analogy to the Fourier expansion of the azimuthal distribution $\frac{dN}{d\phi}$, we have   Fourier expanded azimuth  dependent slope parameter. The expansion coefficients of the slope parameter and that of azimuthal distribution appear to be strongly correlated. Strong correlations between two expansions coefficients suggest that like the azimuthal distributions, azimuth dependent  slope parameter can be used to characterise the momentum anisotropy. 

The paper is organised as follows: in section \ref{sec2} the hydrodynamic model used presently is briefly discussed. In section \ref{sec3} simulation results are discussed and finally in section \ref{sec4} conclusions are drawn.

\section{Hydrodynamic model} \label{sec2}

In event-by-event hydrodynamics, one generally solves for the energy-momentum   and baryon number conservation equations,

\begin{eqnarray} 
\partial_\mu T^{\mu\nu}&=&0, \label{eq2} \\
\partial_\mu J^{\mu}&=&0 \label{eq3}
\end{eqnarray}

\noindent where $T^{\mu\nu}=(\varepsilon+p)u^\mu u^\nu -P g^{\mu\nu}$ is the energy-momentum tensor and $J^\mu=n_B u^\mu$ is the particle 4-current.
$\varepsilon$, $p$, $n_B$ and $u$ are energy density, pressure, net baryon density and  hydrodynamic 4-velocity respectively. $g^{\mu\nu}=diag(1,-1,-1,-1)$ is the metric tensor. Presently we assume that fluid is baryon free and disregard Eq.\ref{eq3}. We also  disregard any dissipative effect. Assuming boost invariance we solve the equations with the code AZHYDRO-KOLKATA  \cite{Chaudhuri:2008sj}   
  in $(\tau=\sqrt{t^2-z^2},x,y,\eta_s=\frac{1}{2}\ln\frac{t+z}{t-z})$ coordinate system. Hydrodynamics equations (Eq.\ref{eq1}) are closed with an equation of state (EoS) $p=p(\varepsilon)$.
Presently, we use an equation of state where the Wuppertal-Budapest \cite{Aoki:2006we,Borsanyi:2010cj} 
lattice simulations for the deconfined phase is smoothly joined at $T=T_c=174$ MeV, with hadronic resonance gas EoS comprising of all the resonances below mass $m_{res}$=2.5 GeV. Details of the EoS can be found in \cite{Roy:2011xt}. 

Solution of hydrodynamic  equations  requires to specify the   fluid energy density  distribution $\varepsilon(x,y)$, velocity distribution $v_x(x,y),v_y(x,y)$ at the initial time.  A freeze-out prescription is also needed to convert the information about fluid energy density and velocity to invariant particle distribution.  We assume that at the initial time  $\tau_i$=0.6 fm  initial fluid velocity is zero, $v_x(x,y)=v_y(x,y)=0$. The freeze-out temperature is fixed at $T_F$=130 MeV. For the initial energy density distribution     we resort to Monte-Carlo Glauber model.   Details of the Monte-Carlo Glauber model can be found in \cite{Alver:2008aq}.
In a Monte-Carlo Glauber model, according to the density distribution of the colliding nuclei,    two nucleons are randomly chosen. They are assumed to interact if the transverse separation  is below $\sqrt\frac{\sigma_{NN}}{\pi}$. For RHIC energy collisions $\sigma_{NN}\approx 42 mb$, a value used in the present simulations.
  Transverse position of the participating nucleons (which are known in each event) will fluctuate from event-to-event. If a particular event has $N_{part}$ participants,  participants positions in the transverse plane can be labeled as, $(x_1,y_1), (x_2,y_2)....(x_{N_{part}},y_{N_{part}})$. Energy density distribution in the particular event can be obtained by assuming that    each participant deposits energy $\varepsilon_0$ in the transverse plane,  

\begin{equation}\label{eq4}
\varepsilon(x,y) \approx \varepsilon_0 \sum_{i=1}^{N_{part}}  \delta(x-x_i,y-y_i)
\end{equation}

For use in fluid dynamical models, the discrete density distribution is smoothed by smearing   the participant positions by a Gaussian function, 
 
\begin{eqnarray} 
&&\varepsilon(x,y)=\varepsilon_0 \sum_{i=1}^{N_{part}}  g_{Gauss}(x-y,x_i,y-y_i,\sigma) \label{eq5}\\
&&g_{Gauss}(x-x_i,y-y_i,\sigma) \propto e^{-\frac{{(x-x_i)^2+(y-y_i)^2}}{2\sigma^2}} \label{eq6},
\end{eqnarray}

In  \cite{Schenke:2011zz}\cite{RihanHaque:2012wp}, influence of the Gaussian width i.e. the smoothing parameter $\sigma$, on flow coefficients were studied in Au+Au collisions. Elliptic and triangular flows are minimally influenced by the smoothing parameter   $\sigma$. Higher flow coefficients, however are influenced by the  choices of $\sigma$. Presently, we have used $\sigma$=0.25 fm.

\section{Results}  \label{sec3}
\subsection{Feasibility of slope parameter as a probe for momentum asymmetry} 

\begin{figure}[t]
\center
\resizebox{0.45\textwidth}{!}{%
\includegraphics{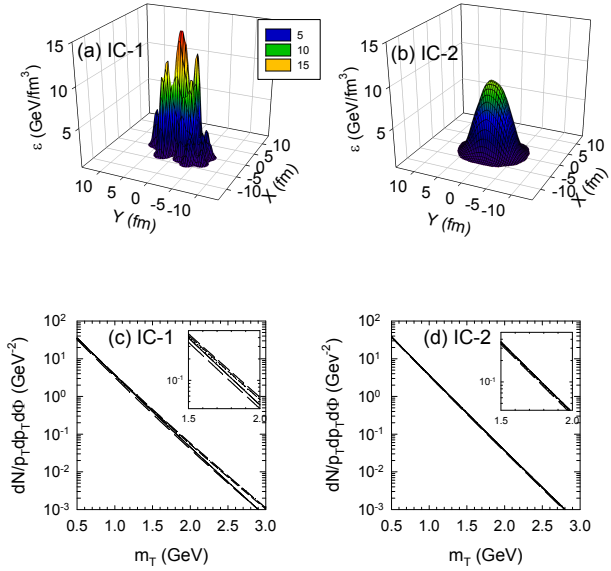} 
}
\caption{(color online) (a) Distribution of initial energy density in a single MC event (IC-1), (b) distribution of energy density averaged over  500 MC events (IC-2). 
(c) Transverse momentum distribution for $\pi^-$ from hydrodynamic simulation of initial
condition IC-1, (d) Transverse momentum distribution for $\pi^-$ from hydrodynamic  simulation of initial
condition IC-2.}
\label{F1}
\end{figure}

Let us first demonstrate the feasibility of slope parameter in characterising momentum  anisotropy of final state particles. We have simulated     30-40\% Au+Au collision at RHIC energy, with two initial condition IC-1 and IC-2. IC-1 is a randomly chosen MC Glauber model initial condition and IC-2 is the initial condition obtained by averaging over average of 500 MC events. In Fig.\ref{F1}a and (b), energy density distributions for the two initial conditions are shown. When averaged over large number of events, the initial condition is rather smooth as it is in smooth hydrodynamics (obtained from optical Glauber model calculations). Initial condition IC-1, corresponding to a single   MC event clearly shows granular structures. 
   
In Fig.\ref{F1}c and d, angular dependence of $\pi^-$ spectra from the evolution of initial   distributions IC-1 and IC-2 are shown. Spectra are shown for azimuthal angle $\phi$=0, 1.26, 2.51, 3.77, 5.03 and 6.28 respectively. In the inset of \ref{F1}c and d, spectra are shown in a limited $p_T$ range.  For IC-2, as expected, the spectra do not show any angular dependence. The angular dependence however is manifestly presented in IC-1. We also note that  
  at any angle the spectrum is of exponential nature, $~exp(-m_T/T)$. Azimuth dependent slope parameter then can be used to probe momentum anisotropy of the final state particles.  
  In smooth hydrodynamics, the azimuth independent slope parameter can be interpreted as effective temperature of a single fireball emitting particles. Azimuth dependent slope parameter then can be interpreted as the temperatures of multiple sources.

\begin{figure}[t]
\center
\resizebox{0.45\textwidth}{!}{%
\includegraphics{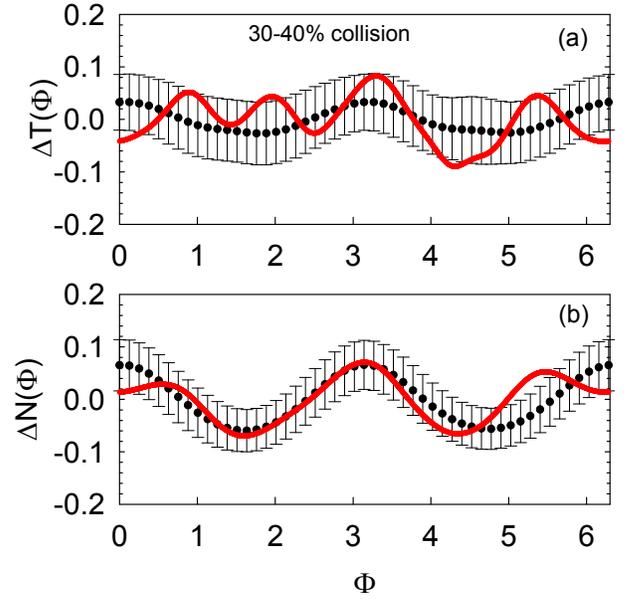} 
}
\caption{(color online) (a) Variation of the fluctuations in the slope parameter $\Delta T(\phi)$ with azimuth. The red line is for a single MC Glauber event and the filled circles with error bars are for average of 500 MC Glauber events. (b) same as in (a) but for fluctuations in the azimuthal distribution $\Delta N$.}
\label{F2}
\end{figure}

\subsection{Azimuthal variation of slope parameter in Event-by-event hydrodynamics}

For 500 MC Glauber events we have simulated $\pi^-$ production in  30-40\% Au+Au collisions at RHIC. For each event we find the slope parameter ($T(\phi)$) between $p_T$=1-2 GeV, and compute the   fluctuations in $T(\phi)$ as, 

\begin{equation}
 \Delta T(\phi)=\frac{T(\phi)-\overline{T}}{\overline{T}}; \hspace{0.5cm} \overline{T}=\frac{\int_0^\infty d\phi T(\phi) }{2\pi}
\end{equation}

In simulations we also have calculated the azimuthal distribution $\frac{dN}{d\phi}$ in each of the event. Fluctuations in azimuthal distribution are calculated in a similar way,

\begin{equation}
 \Delta N(\phi)=\frac{\frac{dN}{d\phi}-\overline{\frac{dN}{d\phi}}}{\overline{\frac{dN}{d\phi}}}; \hspace{0.5cm} \overline{\frac{dN}{d\phi}}=\frac{\int_0^\infty d\phi \frac{dN}{d\phi} }{2\pi}
\end{equation}

In Fig.\ref{F2} azimuthal fluctuations $\Delta T(\phi)$ and $\Delta N(\phi)$ are shown. In Fig.\ref{F2}a, the circles with error bars shows the azimuthal variation for the event averaged fluctuations   $\langle \Delta T(\phi)\rangle $. When averaged over all the events, the slope parameter does not show rapid variation with azimuth. However, large error bars do indicate that at any azimuth, event-by-event variations are rather large. In Fig.\ref{F2}a, the red line is the fluctuations in the event IC-1. $T(\phi)$   more oscillates more rapidly than the event averaged value. Fig.\ref{F2}b shows the azimuthal variation $\Delta N(\phi)$. Event averaged   $\langle \Delta N\rangle$  fluctuations are   similar to that of the slope parameter, though more pronounced. In Fig.\ref{F2}b, the red line indicate azimuthal variation of $\Delta N$ in the event IC-1. In a single event, unlike $\Delta T$, $\Delta N$ oscillates less rapidly.

 \subsection{Fourier expansion of slope parameter}  
 
As in Eq.\ref{eq1}, azimuthal anisotropy of slope parameter can be studied   by Fourier expansion, 

\begin{equation} \label{eq8}
T(\phi)=T_0 \left[ 1+ 2\sum_n T_n cos(n\phi-n\psi_n)\right ], n=1,2,3...
\end{equation} 

\noindent where the expansion coefficients are labeled as $T_1$, $T_2$.. etc. 
In Eq.\ref{eq8} (also in Eq.\ref{eq1}) $\psi_n$ is the participant plane angle. 
In event-by-event hydrodynamics, one generally characterise the asymmetry of the initial collision zone in terms of various moments of the eccentricity ($\epsilon_n$) \cite{Alver:2010gr},\cite{Alver:2010dn},\cite{Teaney:2010vd},

\begin{eqnarray} 
\epsilon_n e^{in\psi_n} &=&-\frac{\int \int \varepsilon(x,y) r^n e^{in\phi}dxdy}{\int \int\varepsilon(x,y) r^n dxdy}, n=1,2,3.. \label{eq9}
\end{eqnarray} 

\noindent which also determine the participant plane angle $\psi_n$.

\begin{figure}[t]
\vspace{1.0cm} 
\center
\resizebox{0.45\textwidth}{!}{%
\includegraphics{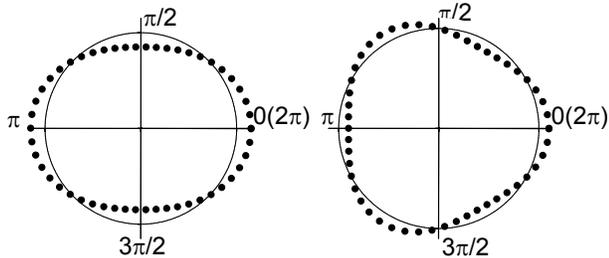} 
}
\caption{(Color online) Pictorial depiction of particle distribution with finite $v_2$ and $v_3$ in the transverse plane. The red lines in the left and right panels show the polar plot of the azimuthal distribution $\frac{dN}{d\phi}$ with finite $v_2$ and finite $v_3$. The black circles shows the distribution when $v_2$=$v_3$=0.}
\label{F3}
\end{figure}

To understand the implications of the coefficients $T_n$, let us remember that the physical implications of the flow coefficients $v_n$ in Eq.\ref{eq1}. In Fig.\ref{F3} schematic diagram of the azimuthal distribution $\frac{dN}{d\phi}$ with finite $v_2$ and with finite $v_3$ are shown in polar coordinates. For finite $v_2$, in the transverse plane, the distribution is elliptical, particles are preferentially produced at $\phi=0$ and $\phi=\pi$. The same for distribution with finite $v_3$ however is triangular in shape. Preferential emission occurs in three direction, $\phi$=0, $2\pi/3$ and $4\pi/3$. In Fig.\ref{F3} solid circles shows the azimuthally symmetric distribution when $v_2$ or $v_3$ are zero. Microscopically, due to multiple collisions, initial asymmetry in spatial density distribution gets converted into asymmetry in the momentum space distribution. 
If it is assumed that particles are emitted from a hot fireball in the momentum space, then for finite $v_2$ ($v_3$),  momentum space density distribution of the fireball, in the transverse plane is elliptical (triangular). The coefficients $v_n$ in Eq.\ref{eq1} are called flow coefficients as they indicate preferential flow of particles in certain directions. 
Physical implication of the coefficients $T_n$ in Fourier expansion of the slope parameter is similar to that of the flow coefficients $v_n$. In a distribution with finite $T_2$, in the transverse plane fireball temperature is not uniform  rather elliptical in shape, more along $\phi$=0 and $\pi$ and less in other directions. Similarly fireball temperature will be of triangular shape in distribution with finite $T_3$. In a fireball picture such angular dependence implies that a number of fireballs participate in particle production. Considering the similarity with  flow coefficients $v_n$,   in the following $T_n$  will be called temperature flow coefficients. Similar to flow coefficients, microscopic origin of   temperature flow coefficients is also the asymmetry in the initial (spatial) density distribution.   

In a hydrodynamic model flow coefficients $v_n$ are response of the initial (spatial) asymmetry of the medium produced in the collisions. 
They are expected to be strongly correlated with initial eccentricity measured $\epsilon_n$. Indeed, in smooth hydrodynamics elliptic flow is strongly correlated with initial eccentricity $\epsilon_2$. In event-by-event hydrodynamics correlation between flow coefficient $v_n$, with initial eccentricity measures $\epsilon_n$ has been studied previously \cite{Gardim:2011xv},\cite{Gardim:2012dc},\cite{arXiv:1104.0650},\cite{Qiu:2012uy}. It was shown that
in event-by-event hydrodynamics, elliptic flow is strongly correlated with initial eccentricity.  Comparatively weak correlation was observed   between triangular flow and initial triangularity. Correlation between higher flow coefficients and asymmetry measures gets even weaker. De-correlation of higher order flows could be understood
as due to nonlinear mixing of modes \cite{Gardim:2011xv}. For example, Gardim et al \cite{Gardim:2011xv}  showed that in order
to correctly predict $v_4$ and $v_5$,  one must take into account nonlinear terms proportional $\epsilon^2$ and $\epsilon_2\epsilon_3$ respectively. 

\begin{table}[h] 
\caption{\label{table1} Correlation measure for (i) flow harmonics ($v_n$) and eccentricity ($\epsilon_n)$, (ii) temperature flow coefficients ($T_n$) and eccentricity ($\epsilon_n)$, and (iii) between temperature flow coefficients ($T_n$) and flow coefficient ($v_n$).}
\begin{ruledtabular} 
  \begin{tabular}{|c|c|c|c|c|c|}\hline
  &\multicolumn{5}{c|}{ $C_{measure}$ } \\ \hline
  & $n=1$ &$n=2$ & $n=3$ & $n=4$& $n=5$ \\ \hline 
($v_n,\epsilon_n$)&0.656 & 0.974 & 0.846 & 0.682 & 0.715 \\ 
($T_n,\epsilon_n$)&0.572 & 0.797 & 0.796 & 0.651 & 0.651 \\ 
($T_n,v_n$)       &0.897 & 0.824 & 0.940 & 0.926 & 0.909 \\ 
\end{tabular}\end{ruledtabular}  
\end{table} 

\begin{figure}[t]
\vspace{1.0cm} 
\center
\resizebox{0.45\textwidth}{!}{%
\includegraphics{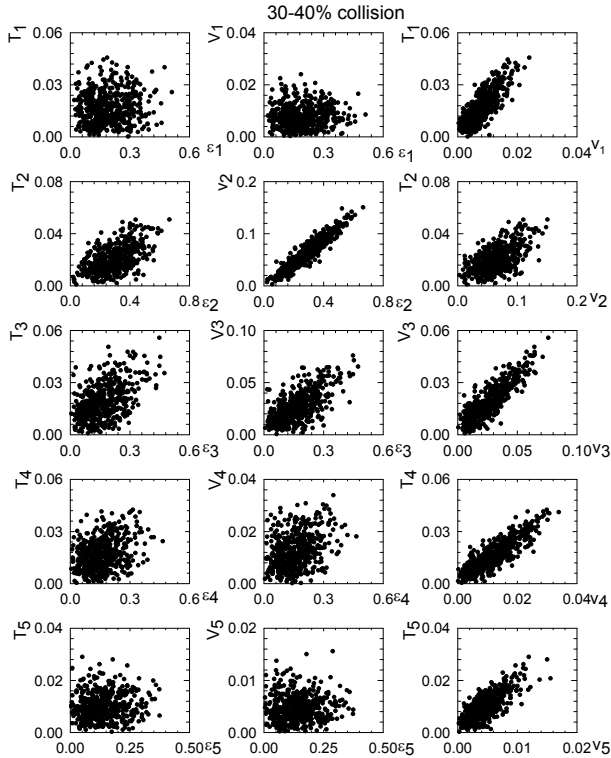} 
}
\caption{Each dot in the plots represent event-by-event hydrodynamic simulation for flow coefficients $v_n$, temperature flow coefficients $T_n$. Left panel: Correlation between temperature flow coefficients $T_n$  with initial eccentricity measure $\epsilon_n$. Middle panel: Correlation between flow coefficients $v_n$ with initial eccentricity measure $\epsilon_n$. Right panel:  Correlation between temperature flow coefficients $T_n$ with flow coefficients $v_n$.  }
\label{F4}
\end{figure}

As we have argued here, temperature flow coefficients $T_n$ are also responses of the initial (spatial) asymmetry of the system. How they are correlated with initial eccentricity measures is an interesting study.
In Fig.\ref{F4}, we have studied the correlation between initial eccentricity ($\epsilon_n$) and temperature flow coefficients ($T_n$). For each of the 500 MC-Glauber events the slope parameter $T(\phi)$ was Fourier expanded to find out the temperature flow coefficients $T_n$, n=1,2,..5. In the left panel of Fig.\ref{F4}, for the 500 events, the temperature flow coefficients  $T_n$ (n=1,2,..5) are plotted against initial eccentricity measure $\epsilon_n$.  For a perfect correlation $T_n\propto \epsilon_n$   and all the events should lie on a straight line. It is evident from our simulations that temperature flow coefficients are weakly correlated with initial eccentricity measures.

The result is not a complete surprise. From our previous studies, we know that with the exception of elliptic flow $v_2$, the flow coefficient $v_n$ are also weakly correlated with initial eccentricity measures. It is evident from the middle panel of Fig.\ref{F4} where  we have plotted the flow coefficients $v_n$ against the eccentricity measures $\epsilon_n$. As expected, only $v_2$ appears to be strongly correlated with $\epsilon_2$. Other flow coefficients are weakly correlated with eccentricity measures. In the right panel of Fig.\ref{F4}, we have plotted temperature flow coefficients $T_n$ against the flow coefficient $v_n$. Interestingly, temperature	flow coefficients $T_n$ and flow coefficient $v_n$ appear to be strongly correlated.

In  \cite{Chaudhuri:2011pa}\cite{Chaudhuri:2012wn} a quantitative measure was defined to quantify the correlation between flow coefficients and initial spatial asymmetry measure.

\begin{equation} \label{eq11}
C_{measure}(n)=1-\frac{\sum_i [ v_n^i(\epsilon_n) -v_{n,st.line}(\epsilon_n) ]^2}{\sum_i [ v^i_{random}(\epsilon) -v_{st.line}(\epsilon) ]^2}
\end{equation}

$C_{measure}$ 
  essentially measures the dispersion of the simulated flow coefficients from the best fitted straight line, relative to completely random flow coefficients. It varies between 0 and 1. 
If flow coefficients are perfectly correlated then $v_n \propto \epsilon_n$ and $C_{measure}$ is identically unity. For completely random flow coefficients, $C_{measure}$=0. To obtain an even ground for comparison of $C_{measure}$ for different flow coefficients, the flow coefficients ($v_n$) and the asymmetry parameters ($\epsilon_n$) are scaled to vary between 0 and 1. A similar equation can be used to   quantify the correlation between temperature flow coefficients $T_n$ and initial asymmetry measure $\epsilon_n$ or between temperature flow coefficients $T_n$ and flow coefficients $v_n$.    In table.\ref{table1}, we have listed the $C_{measure}$ values for correlation between $v_n$ and $\epsilon_n$, between $T_n$ and $\epsilon_n$ and between $T_n$ and $v_n$. Quantitatively  compared to flow coefficients $v_n$, temperature flow coefficients $T_n$ are less correlated with $\epsilon_n$. The difference is most prominent for second harmonic n=2 and lessened in higher order harmonics. $C_{measure}$ values also indicate that the correlation between temperature flow coefficients $T_n$ and flow coefficient $v_n$ are rather high for all n. $T_n$ and $v_n$ are much better correlated than the correlation between $v_n$ and $\epsilon_n$ or between $T_n$ and $\epsilon_n$.  

Event-by-event hydrodynamic simulation result that the temperature flow coefficients $T_n$ and flow coefficients $v_n$ are rather strongly correlated strongly suggests that apart from the azimuthal distribution $\frac{dN}{d\phi}$, the azimuthal fluctuations in the slope parameter of invariant spectra can as well be used to study final state momentum anisotropy. The fluctuations can be measured experimentally and can serve as additional constrains for    hydrodynamic modeling.

\section{Summary and conclusions}\label{sec4}

In event-by-event hydrodynamical model for heavy ion collisions, due to fluctuations in participant positions, the initial conditions fluctuates event-by-event. These fluctuations are manifested as momentum anisotropy in the final particle distribution and in general are studied by Fourier expanding the azimuthal distribution, the expansion coefficients characterising the momentum anisotropy. By explicit event-by-event hydrodynamic simulation we have shown that in event-by-event hydrodynamics, the slope parameter of the invariant distribution will also fluctuates with azimuth and can as well  be used to study the momentum anisotropy.   
For 500 MC Glauber events, we have simulated Au+Au collisions at RHIC and from the invariant pion spectra computed the azimuthal distribution $\frac{dN}{d\phi}$ and from an exponential fit to the spectra between $p_T$=1-2 GeV, also the slope parameter $T(\phi)$.  $\frac{dN}{d\phi}$ and $T(\phi)$ were Fourier expanded   and studied its correlation between expansion coefficients and initial eccentricity measures. With the exception of n=2, expansion coefficients $v_n$ for the azimuthal distribution $\frac{dN}{d\phi}$ are weakly correlated with the initial eccentricity measures $\epsilon_n$. Without any exception, the expansion coefficients $T_n$ for the slope parameter  are also weakly correlated with the initial eccentricity measures $\epsilon_n$.
The coefficients $v_n$ and $T_n$ are however strongly correlated. Strong correlation between $v_n$ and $T_n$ suggest that they complement each other. If  measured experimentally fluctuations in slope parameter can be used to study momentum anisotropy and can serve as additional constraint for hydrodynamical modeling.

\end{document}